\documentclass[10pt, conference, compsocconf]{IEEEtran}

\usepackage{cite}
\usepackage{amsmath,amssymb,amsfonts}
\usepackage[linesnumbered,ruled,vlined]{algorithm2e}
\usepackage{graphicx}
\usepackage{textcomp}
\usepackage{bm}
\usepackage{theorem}
\usepackage{url}
\usepackage{afterpage}
\usepackage{subfig}
\usepackage{wrapfig}
\usepackage{float}
\usepackage{multirow}
\usepackage{makeidx}
\usepackage{listings}
\usepackage{color,lstcoq,url}
\usepackage{dirtree}
\usepackage{verbatim}

\usepackage{article}

\definecolor{ltblue}{rgb}{0,0.4,0.4}
\definecolor{dkblue}{rgb}{0,0,0}
\definecolor{dkgreen}{rgb}{0,0.35,0}
\definecolor{dkviolet}{rgb}{1,0,0}
\definecolor{dkred}{rgb}{0,0,0}

\newtheorem{theo}{Theorem}

\newtheorem{lemm}{Lemma}
\newtheorem{example}{Example}

\newcommand{\Enc}{\mathsf{E}}

\def\BibTeX{{\rm B\kern-.05em{\sc i\kern-.025em b}\kern-.08em
    T\kern-.1667em\lower.7ex\hbox{E}\kern-.125emX}}

\begin{document}

%

\onecolumn

\noindent
{\LARGE IEEE Copyright Notice}

\vspace{2em}

\noindent
Copyright (c) 2020 IEEE.

\noindent
Personal use of this material is permitted. Permission from IEEE must be obtained for all other uses, in any current or future media, including reprinting/republishing this material for advertising or promotional purposes, creating new collective works, for resale or redistribution to servers or lists, or reuse of any copyrighted component of this work in other works.

\vspace{3em}

\noindent
Accepted to be published in the 14th International Symposium on Theoretical Aspects of Software Engineering (TASE), 2020.

\newpage
\twocolumn


\title{
Formalizing the Soundness of the Encoding Methods of SAT-based Model Checking
}

\author{\IEEEauthorblockN{Daisuke Ishii}
\IEEEauthorblockA{School of Information Science \\
JAIST \\
Ishikawa, Japan \\
Email: {dsksh@acm.org}}
\and
\IEEEauthorblockN{Saito Fujii}
\IEEEauthorblockA{
DENSO TEN Limited \\
Hyogo, Japan \\
Email: {donn2005s@gmail.com}}
}

\maketitle

\begin{abstract}
    One of the effective model checking methods is to utilize the efficient decision procedure of SAT (or SMT) solvers.
    In a SAT-based model checking, a system and its property are encoded into a set of logic formulas and the safety is checked based on the satisfiability of the formulas.
    As the encoding methods are improved and crafted (e.g., $k$-induction and IC3/PDR), verifying their correctness becomes more important.
    %
    This research aims at a formal verification of the SMC methods using the Coq proof assistant.
    Our contributions are twofold:
    (1) We specify the basic encoding methods, $k$-induction and (a simplified version of) IC3/PDR in Coq as a set of simple and modular encoding predicates.
    (2) 
    We provide a formal proof of the soundness of the encoding methods
    based on our formalized lemmas on state sequences and paths.
    %
    The specification of the SMC methods and the soundness proofs are available at \url{https://github.com/dsksh/coq-smc/}.
\end{abstract}


\section{Introduction}

SAT-based model checking (SMC, Sect.~\ref{s:smc}) is a symbolic model checking method that delegates the main reasoning/search process to efficient SAT (or SMT) solvers.
An essential process in SMC is to encode a property of a state transition system into a propositional (or predicate) logic formula. Checking the satisfiability of the encoded formula entails the safety of the target system.
Various SMC methods have been proposed, e.g., \cite{Biere1999,Sheeran2000,McMillan2003,Bradley2011,Een2011,Claessen2012}, each of which is carefully designed by e.g. unrolling of execution paths, inductions on paths, and over-approximation of states (Sect.~\ref{s:smc} and \ref{s:smc:unbounded}).

Formal verification is a way to provide a reliable SMC tool.
To the authors' knowledge, there does not exist a verified SMC tool.
A verification involves proving the correctness of the encoding methods, which are based on various properties on states and execution paths of a transition system.
Formalizing the correctness and the underlying properties using a proof assistant is not a trivial task.

This research aims at formal verification of the SMC methods using the Coq proof assistant. 
We first formalize the transition systems, properties and several SMC methods with Coq (Sect.~\ref{s:impl}).
Then, we verify that the methods correctly encode the safety of the target system; 
in other words, we formalize the soundness proof of each SMC algorithm (Sect.~\ref{s:v}).
%
We also investigate a formalization of the properties on states and paths of a system, 
which are considered explicitly in SMC. 
Our contributions are summarized as follows:
\begin{itemize}
\item We formalize the SMC methods in a simplified manner that 
    adapts the existing methods into a generic scheme described in Sect.~\ref{s:smc} and \ref{s:smc:unbounded};
    each SMC method is presented as an implementation of the encoding methods,
    which are then formalized by a shallow embedding into Coq (Sect.~\ref{s:impl:spec}).
    Also, we demonstrate that our Coq specification works as a prototype SMC tool (Sect.~\ref{s:impl:smc}).
%
\item We formally verify the soundness of the encoding methods; in particular, for the forward, backward, $k$-induction, and IC3/PDR methods (Sect.~\ref{s:v}).
    In the formalization, we explicitly specify the execution paths of a system.
    To facilitate this task, we develop a small yet useful theory of state sequences and paths, which contains lemmas on various properties and operations.
    Sect.~\ref{s:res} discusses some characteristics of our verification results.
\end{itemize}

%
We consider that our result will help to provide a reliable SMC tool and 
serve as a formal scheme in the development of new encoding methods.

\section{SAT-based Model Checking}
\label{s:smc}

This section introduces basics of SMC (Sect.~\ref{s:smc:scheme}) and an encoding process for bounded safety (Sect.~\ref{s:smc:bnd}).


We consider a 
set of \emph{states}, each of which is typically encoded as a fixed-length bit vector or an integer; 
$S$ denotes a set of \emph{states}.
A \emph{state transition system} is specified by a pair $(I,T)$ of an \emph{initial condition} $I \subseteq S$ and a \emph{transition relation} $T \subseteq S \times S$.
An \emph{initial path} of a system is a sequence of states $s_0 s_1 \cdots$ such that $s_0 \in I$ and $T(s_i, s_{i+1})$ for $i \geq 0$.
In the following, we denote the path by $s_{[0..]}$ and a path fragment $s_i \cdots s_j$ by $s_{[i..j]}$.
We say a state $s_i \in S$ is \emph{reachable} iff there exists an initial path fragment $s_{[0..i]}$. 
Given a \emph{property} $P \subseteq S$, we say a system $(I,T)$ is \emph{safe} iff every reachable state satisfies $P$;
the \emph{safety} is defined as
\begin{gather}
    \mathit{safety}(I, T, P) ~:\leftrightarrow~ \nonumber \\
    \label{eq:safety}
    \forall i \!\in\! \mathbb{N},\ \forall s_{[0..i]},~ I(s_0) \ \to\ 
    \mathit{path}_T(s_{[0..i]}) \ \to\ P(s_i), 
\end{gather}
where
\begin{gather}
    \label{eq:path}
    \mathit{path}_T(s_{[o..i]}) :\leftrightarrow
    \begin{cases}
        \bigwedge_{o \leq j < i} T(s_j,s_{j+1}) & \text{if $o<i$}, \\
        \mathrm{true}                   & \text{if $o=i$}.
    \end{cases}
\end{gather}

\begin{example}
    \label{ex:system}
    We can consider a flawed 3-bit shift register~\cite{Biere1999} as a simple state transition system with {$I(s) :\leftrightarrow s \!\geq\! 4$} and $T(s,s') :\leftrightarrow (2s+1)\mod 8 = s'$,
    where bits are interpreted as decimal numbers.
    It is safe with the property $P(s) :\leftrightarrow s \neq 0$.
    An example initial path is $100_2;001_2;011_2;111_2;111_2;\cdots$.
\end{example}

\emph{SAT-based model checking (SMC)}~
\cite{Biere1999,Biere2018} 
is a formal verification method for transition systems, which exploits the efficient decision procedure of SAT (propositional satisfiability) or SMT (satisfiability modulo theories) solvers.
In a model checking process, a bound $k \!\geq\! 0$ is given to restrict the depth of the search space.
As a result, either (R1) a safety proof of the system, (R2) partial safety proof bounded by parameter $k$ or (R3) a counterexample of length $k$ or less will be obtained.
%
Various SMC methods have been proposed, e.g. \cite{Sheeran2000,McMillan2003,Bradley2011,Een2011,Claessen2012}.
There exist a number of SMC tool implementations, e.g.,
nuXmv~\cite{Bozzano2019} and Kind 2~\cite{Mebsout2016}.

\subsection{Generic Scheme of SMC}
\label{s:smc:scheme}

\begin{algorithm}[tb]
    \SetAlgoLined
    \SetKwInOut{Input}{Input}\SetKwInOut{Output}{Output}
    \Input{$I$, $T$, $P$, $k$}
    \Output{$\True$ or $\False$}
    \BlankLine
    $F$ := $\Enc_\bullet(I,T,P,k)$\;
    \For{$(\forall s_{[0..m]}, f_i) \in F$}{
        $F$ := $F \{ (\forall s_{[0..m]}, f_i) \mapsto \neg\mathsf{CheckSat}(\neg f_i) \}$\;
    }
    \Return{$\mathsf{Decide}(F)$}\;
    \caption{A generic SMC scheme.}
    \label{a:bmc}
\end{algorithm}


\begin{algorithm}[tb]
    \SetAlgoLined
    \SetKwInOut{Input}{Input}\SetKwInOut{Output}{Output}
    \Input{$I$, $T$, $P$}
    \Output{$\True$ or $\False$}
    \BlankLine
    \For{$k \geq 0$}{
        \uIf{$\mathsf{Alg1}_{\True}(I,T,P,k)$}{
            \Return{$\True$}\;
        }
        \ElseIf{$\neg \mathsf{Alg1}_{\False}(I,T,P,k)$}{
            \Return{$\False$}\;
        }
    }
    \caption{Iterative process for unbounded MC.}
    \label{a:it}
\end{algorithm}

Given a target system $(I,T)$, a property $P$ and a bound $k$, an SMC procedure \emph{encodes} the safety of the system into a logic formula $F$
and checks its satisfiability using a SAT/SMT solver.
This procedure can be summarized as Alg.~\ref{a:bmc}.
Each SMC method provides their own encoding method; therefore, in the algorithm, it is parameterized as $\Enc_\bullet$ at Line~1.
An encoding method generates a formula $F$ that describes a given system and a property by considering only a finite number (which depends on $k$) of states.
As a result, multiple validity problems of the form $\forall{s_{[0..m]}}, {f_i}$ are generated as sub-formulas in $F$; then, they are discharged using a SAT/SMT solver, and those sub-formulas are substituted with the results, i.e. true or false (Line~3).
Here, 
we denote the substitution of a sub-formula $G$ in $F$ with $H$ by $F\{G \mapsto H\}$;
the validity problem is solved by checking the unsatisfiability of the negation of $f_i$.
An encoded formula $F$ can be an arbitrary logical combination of the sub-formulas, e.g., $\neg(\forall{s_{[0..m]}}, {f_1}) \land (\forall{s_{[0..m]}}, {f_2}) \lor \cdots$.
At Line~5, the validity of $F$ is determined, typically with a case analysis.

\subsection{Encoding Bounded Safety}
\label{s:smc:bnd}

As a basic encoding method, we consider to encode a bounded safety property.
For SMC with the bound $k$, 
we prepare $k\!+\!1$ state variables $s_{[0..k]}$. 
Then, the following predicate formula states that every sequence $s_{[0..i]}$, which represents an initial path fragment of length $k$ or less, reaches a safe state.
\begin{gather} 
    \nonumber
    \Enc_{\eqref{eq:ksafety}}(I,T,P,k) ~:\leftrightarrow~ \\
    \label{eq:ksafety}
    \bigwedge_{0 \leq i \leq k} \!
    \bigr(\forall s_{[0..i]},~
    I(s_0) \to \mathit{path}_T(s_{[0..i]}) \to P(s_i)
    \bigr). 
\end{gather} 
In the above, $I$, $T$ and $P$ are predicates on states that specify the target system model and the property.
By running Alg.~\ref{a:bmc} with $\Enc_{\eqref{eq:ksafety}}$, which is supported by a SAT or SMT solver,
we are able to verify whether the system is safe within the bound $k$ (R2) or there exists a counterexample (R3).

To use a SAT solver as the $\mathsf{CheckSat}$ procedure in Alg~\ref{a:bmc}, the body of an encoded formula has to be (the negation of) a conjunction.
Hence, the formula~\eqref{eq:ksafety} should be modified as
\begin{equation}
    \label{eq:ksafety:cnf}
    \bigwedge_{0 \leq i \leq k} \!
    \bigr( \forall s_{[0..i]},~
    \neg(\ I(s_0) \,\land\, \mathit{path}_T(s_{[0..i]}) \,\land\, \neg P(s_i))
     \bigr),
\end{equation} 
which is equivalent to \eqref{eq:ksafety} in classical logic.

\section{Unbounded Model Checking}
\label{s:smc:unbounded}

For a SAT-based unbounded model checking, 
which concerns a safety for arbitrary lenghs of path fragments (R1),
we typically perform an iteration of encoding and satisfiability checking for $k \geq 0$.
This process is illustrated as Alg.~\ref{a:it},
which contains two calls for Alg.~\ref{a:bmc} (Lines~2 and 4) with different encoding methods for checking safety (R1) or unsafety (R3).
Whenever the calls result in $\True$ or $\False$, respectively, the process terminates.
The termination 
depends on the completeness of the encoding methods.

This section describes the dedicated encoding processes, $k$-induction (Sect.~\ref{s:smc:enc}) and IC3/PDR (Sect.~\ref{s:smc:pdr}). 

\subsection{Encoding Unbounded Safety}
\label{s:smc:enc}

Encoding methods for unbounded safety that assume a set of lasso-shaped paths and apply an induction have been proposed.
Sheeran et al.\cite{Sheeran2000} have proposed six algorithms for this purpose,
which are intended for an efficient encoding and reasoning of lasso-shaped paths.
Some of the algorithms also utilize the $k$-induction principle for an efficient SMC.
%

First, we prepare a shorthand to describe that $s_{[o..k]}$ is a loop-free path.
\begin{align} 
  \label{eq:loopfree}
  \mathit{loopF}_T(s_{[o..k]}) \ :\leftrightarrow\  \mathit{path}_T(s_{[o..k]}) \land\! \bigwedge_{o \leq i <  j \leq k} s_i \neq s_j 
\end{align} 
Then, a sufficient condition for the safety (R1) can be encoded in two ways.
\begin{multline}
    \label{eq:forward}
    \Enc_{\eqref{eq:forward}}(I,T,P,k) ~:\leftrightarrow~
    \Enc_{\eqref{eq:ksafety}}(I,T,P,k) ~\land
    \\
    \bigl( \forall s_{[0..k]},~ I(s_0) \,\to\,
    \neg\mathit{loopF}_T(s_{[0..k]}) \bigr)
\end{multline}
\vspace{-2em}
\begin{multline}
    \label{eq:backward}
    \Enc_{\eqref{eq:backward}}(I,T,P,k) ~:\leftrightarrow~
    \Enc_{\eqref{eq:ksafety}}(I,T,P,k) ~\land 
    \\
    \bigl(\forall s_{[0..k]},~ \mathit{loopF}_T(s_{[0..k]}) \,\to\,
    P(s_{k}) \bigr)
\end{multline}
Either of the encoding methods $\Enc_\eqref{eq:forward}$ and $\Enc_\eqref{eq:backward}$ considers a fixed point of the reachability analysis from the initial states or the unsafe states, respectively.
The second quantified sub-formula of $\Enc_\eqref{eq:forward}$ checks that all the initial paths of length $k$ contain a loop.
The implication sub-formula of $\Enc_\eqref{eq:backward}$ is equivalent to $\neg P(s_{k}) \to \neg \mathit{loopF}_T(s_{[0..k]})$ in classical logic; it checks that all paths of length $k$ from an unsafe state contain a loop.
Since the safety within the bound $k$ is checked with $\Enc_\eqref{eq:ksafety}$, the safety R1 is inferred as a result.
%
%
As for $\Enc_{\eqref{eq:ksafety}}$, we can modify the second parts of $\Enc_{\eqref{eq:forward}}$ and $\Enc_{\eqref{eq:backward}}$ as follows to have the sub-formulas of a conjunctive form.
\begin{gather}
    \label{eq:forward:cnf}
    \forall s_{[0..k]},~ 
    \neg( I(s_0) \,\land\, \mathit{loopF}_T(s_{[0..k]}) ) \\
    \label{eq:backward:cnf}
    \forall s_{[0..k]},~ 
    \neg( \mathit{loopF}_T(s_{[0..k]}) \,\land\, \neg P(s_{k}) ) 
\end{gather}

\begin{example}
    \label{ex:fb}
    For the system in Ex.~\ref{ex:system}, $\Enc_{\eqref{eq:forward}}$ holds iff $k \!\geq\! 4$ since all initial paths are safe and they reach the invariant state $111_2$ before the fourth step.
    $\Enc_{\eqref{eq:backward}}$ holds iff {$k \!\geq\! 1$} since the unsafe state $000_2$ does not have a predecessor state.
\end{example}

Sheeran et al.'s first algorithm~\cite{Sheeran2000} is a hybrid method of the two, whose encoding method is equivalent to $\Enc_{\eqref{eq:forward}} \lor \Enc_{\eqref{eq:backward}}$.
Sheeran et al.'s subsequent algorithms apply a $k$-induction and encode the safety as follows.
\begin{align}
  \label{eq:kinduction}
  & \Enc_{\eqref{eq:kinduction}}(I,T,P,k) ~:\leftrightarrow~ 
  \Enc_{\eqref{eq:ksafety}}(I,T,P,k) ~\land
  \\
  & \Bigl(\forall s_{[0..(k+1)]}, \mathit{path}_T(s_{[0..(k+1)]}) \!\to\!\! \bigwedge_{0 \leq i \leq k}\!\!P(s_i) \!\to\! P(s_{k+1})\Bigr) \nonumber
\end{align}

To have an unbounded model checker using our scheme of Alg.~\ref{a:it},
$\Enc_\bullet$ in $\mathsf{Alg1}_{\True}$ can be implemented as either $\Enc_{\eqref{eq:forward}}$, $\Enc_{\eqref{eq:backward}}$, $\Enc_{\eqref{eq:kinduction}}$ or $\Enc_{\eqref{eq:forward}} \lor \Enc_{\eqref{eq:backward}}$, and it can be implemented as $\Enc_{\eqref{eq:ksafety}}$ in $\mathsf{Alg1}_{\False}$.

\subsection{IC3/PDR Method}
\label{s:smc:pdr}

Bradley~\cite{Bradley2011} has proposed an SMC method called IC3 or PDR (property directed reachability)~\cite{Een2011} (we refer to the method as PDR in the sequel); 
it handles a sequence of \emph{over-approximations} that encloses path fragments with sets of clauses; 
also, it uses the \emph{inductive relative} relation between states to refine an over-approximation.
In this paper, we do not go into the details of the method but focus on the post-conditions of the PDR method, given an over-approximation.
We again adapt the method into our scheme of Alg.~\ref{a:bmc} and Alg.~\ref{a:it}.

\begin{figure}[!t]
    \centering
    \includegraphics[width=.35\textwidth]{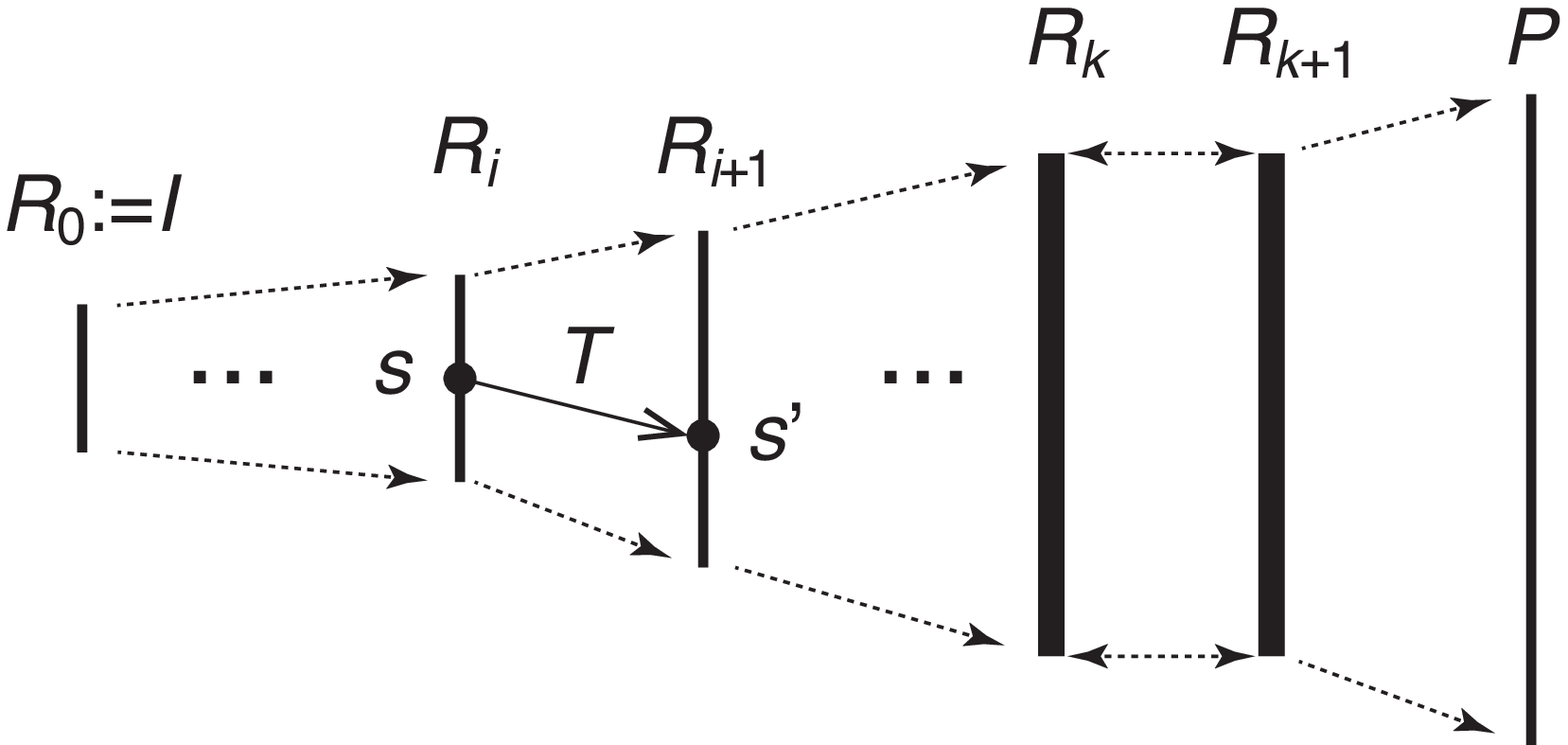} 
    \caption{Inductive strengthening of $P$.}
    \label{f:pdr}
\end{figure}

Assume a sequence $R_0 R_1 \cdots$ (denoted by $R_{[0..]}$ below) of over-approximations of a set of reachable states in $S$.
Then, the post-condition of PDR for the true case is described as follows (we refer to the whole formula as \eqref{eq:pdr:t} and to each sub-formula as (a)--(e)).
\begin{align}
    \Enc_{\eqref{eq:pdr:t}}(I,T,P,k) \ :\leftrightarrow\ &
    \exists R_{[0..]},
    \label{eq:pdr:t:a}
    \bigl( \forall i, \forall s,~ I(s) \to R_i(s) \bigr) \land \tag{a} \\
    \label{eq:pdr:t:b}
    & \bigl( \forall i, \forall s,~ R_i(s) \to P(s) \bigr)\ \land \tag{b} \\
    \label{eq:pdr:t:c}
    & \bigl( \forall i, \forall s,~ R_i(s) \to R_{i+1}(s) \bigr)\ \land \tag{c} \\
    \label{eq:pdr:t:d}
    & \hspace*{-7em} \bigl( \forall i \leq k, ~ \forall s,s',~ R_i(s) \land T(s,s') \to R_{i+1}(s') \bigr)\ \land \tag{d} \\
    \label{eq:pdr:t}
    & \bigl( \forall s,~ R_k(s) \leftrightarrow R_{k+1}(s) \bigr) 
    \hspace{1.5em} \text{(e)}
\end{align}
The original algorithm initializes over-approximations as $R_0 \!:\leftrightarrow\! I$ and $R_i \!:\leftrightarrow\! P$ for $i \!\geq\! 1$, and
incrementally refines $R_{[0..(k+1)]}$ (for $k\!:=\!0,\ldots$).
The algorithm repeatedly 
(i) detects a bad state $\tilde{s}$ such that $\exists \tilde{s}', R_k(\tilde{s}) \land T(\tilde{s},\tilde{s}') \land \neg P(\tilde{s}')$,
(ii) generalizes $\neg\tilde{s}$ to a clause $C$
such that $\forall s,s', R_i(s) \land C(s) \land T(s,s') \to C(s')$ (for some $i \leq k$), and 
(iii) refines $R_{[0..(i+1)]}$ by adding $C$ into the over-approximations.
Finally, the algorithm terminates when the sub-formula (e) holds; the resulting over-approximation $R_{k}$ is an \emph{inductive strengthening} of $P$, which is depicted in Fig.~\ref{f:pdr}.

Note that the above post-condition is slightly modified from that of the original algorithm for the sake of simplicity of the verification process.
First, although the original algorithm maintains a set of clauses for each $R_i$, we do not distinguish the set and $R_i$.
Second, we weaken the sub-formula (e) from
$\exists i \leq k, \forall s,~ R_i(s) \leftrightarrow R_{i+1}(s)$.
%

\begin{example}
    \label{ex:pdr}
    For the system in Ex.~\ref{ex:system}, we can have a set of over-approximations specified as $R_0 :\leftrightarrow I$ and $R_i(s) :\leftrightarrow s \geq 1$ for $i \geq 1$.
    It satisfies $\Enc_\eqref{eq:pdr:t}$ iff $k \geq 1$.
\end{example}

The post-condition for the false case, which is extracted from the PDR algorithm, is described below (the sub-formulas are referred to as \eqref{eq:pdr:t:a}--\eqref{eq:pdr:t:c}).
\begin{align}
    \label{eq:pdr:f:a}
    & \Enc_{\eqref{eq:pdr:f}}(I,T,P,k) ~:\leftrightarrow~
    %
    \bigl( \forall s,~ I(s) \to P(s) \bigr) \ \land \tag{a} \\
    \label{eq:pdr:f:b}
    & \bigl( \forall s, s',~ I(s) \to T(s, s') \to P(s')) \bigr) \ \land \tag{b} \\
    \label{eq:pdr:f}
    & \Bigl( \forall R_{[0..(k-1)]}, s_{[0..k]},~ \\
    \label{eq:pdr:f:c}
    & \hspace{0em} I(s_0) \land \bigwedge_{0 \leq i < k} \bigl( R_i(s_i) \land T(s_i,s_{i+1}) \bigr) \to P(s_{k}) \Bigr) \tag{c}
\end{align}
The sub-formulas \eqref{eq:pdr:t:a} and \eqref{eq:pdr:t:b} are for detecting length 0 and 1 counterexamples.
The sub-formula \eqref{eq:pdr:t:c} describes initial and bounded safe paths of length $k$ or less, which is related with $R_{[0..(k-1)]}$.

\section{Specification of SMC Encoding Methods}
\label{s:impl}

In this work, we have realized the SMC scheme in Fig.~\ref{a:bmc} in the Coq proof assistant (Sect.~\ref{s:impl:coq}).
We first explain how we specify 
MC problems and
encoding methods $\Enc_\bullet$ in Coq
(Sect.~\ref{s:impl:spec}).
%
Next, we show that an SMC is performed as a theorem proving in Coq (Sect.~\ref{s:impl:smc}).

\subsection{Coq Proof Assistant and coq2smt}
\label{s:impl:coq}

\emph{Coq}\footnote{\url{https://coq.inria.fr}} (version 8.6.1) is a proof assistant based on the typed lambda calculus, 
which supports predicate logic formulas, algebraic types, higher-order functions, etc. to describe theorems and proofs.
Using the \emph{tactic} mechanism, a proof can be described/performed efficiently and simply. 
In a part of the verification, we rely on the excluded middle axiom of the \verb|Classical_Prop| module (see Sect.~\ref{s:res:cl}).

A plug-in software for Coq, \emph{coq2smt}\footnote{\url{https://github.com/wangjwchn/coq2smt}} (commit \verb|604f72a|), has developed to invoke SMT solvers within a Coq proof via the tactic \verb|smt solve|.
The tactic handles goals of a quantifier-free form involving boolean connectives, equalities, values/variables of type \verb|Z|, etc.; 
it discharges a goal using an SMT solver (e.g. CVC4 and Z3) and reflects the result in the proof context of Coq.

\subsection{State Transition Systems and Encoding Predicates}
\label{s:impl:spec}

To specify a state transition system $(I, T)$, its property $P$ and an over-approximation $R$, we assume the following types based on the definitions in Sect.~\ref{s:smc}.
\begin{lstlisting}[language=Coq, numbers=none, %xleftmargin=10pt
xleftmargin=7pt]
Definition state : Type := Z.
Definition prop: Type := state -> Prop.
Definition trans : Type := state -> state -> Prop.
\end{lstlisting}
We specify the type \verb|state| as \verb|Z| to demonstrate a model checking using SMT solvers.
To pursue the proof tasks in Sect.~\ref{s:v}, any type with equality \verb|=| can be used as \verb|state|.
$I$, $P$ and $R$ are supposed to be typed as \verb|prop|.
For state sequences, we assume the following type.
\begin{lstlisting}[language=Coq, numbers=none, xleftmargin=10pt]
Definition sseq : Type := nat -> Z.
\end{lstlisting}
Here, we have not drawn on existing abstract types (e.g. \verb|list|) because, in our proof scheme for SMC (Sect.~\ref{s:impl:smc}), we intend to expand a proof goal into a flat formula, instead of an inductive reduction.
For instance, the predicate $\mathit{path}$ in Eq.~\eqref{eq:path}, which requires a sequence $\texttt{ss}_{[\texttt{o}..(\texttt{len}-\texttt{o})]}$ to be a path fragment, is specified recursively as follows. Indeed, this enables an expansion with an \verb|unfold| application given specific argument values.
\begin{lstlisting}[language=Coq, numbers=none, xleftmargin=10pt]
Fixpoint path (T : trans) (ss : sseq) (o len : nat) 
: Prop :=
  match len with
  | O => True
  | S len' => path T ss o len' /\ 
                T (ss (o+len')) (ss (o+len))
  end.
\end{lstlisting}

Based on these types, for example, the method $\Enc_\eqref{eq:forward}$ is simply specified as the following predicate:
\begin{lstlisting}[language=Coq, numbers=none, xleftmargin=10pt]
Definition forward_safety (I : prop) (T : trans) 
(P : prop) (k : nat) : Prop :=
  safety_k I T P k /\ lasso_fwd I T k.
\end{lstlisting}
The predicates are defined separately for each sub-formula in Eq.~\eqref{eq:forward};
e.g., the predicate \verb|lasso_fwd| is specified as
%
\begin{lstlisting}[language=Coq, numbers=none, xleftmargin=10pt]
Definition lasso_fwd (I : prop) (T : trans) (k : nat) 
: Prop :=
  forall ss : sseq, ~ (I (ss 0) /\ loop_free T ss 0 k).
\end{lstlisting}
In this way, these predicates are able to be shared between different encoding methods.
For example, a hybrid method of $\Enc_\eqref{eq:forward}$ and $\Enc_\eqref{eq:backward}$ is specified as follows.
\begin{lstlisting}[language=Coq, numbers=none, xleftmargin=10pt]
Definition sheeran1_safety (I : prop) (T : trans) 
(P : prop) (k : nat) : Prop :=
  safety_k I T P k /\ 
  (lasso_fwd I T k \/ lasso_bwd T P k).
\end{lstlisting}

A set of over-approximations (i.e. a sequence of \verb|prop| values) handled by PDR is typed as the type \verb|spseq|.
\begin{lstlisting}[language=Coq, numbers=none, xleftmargin=10pt]
Definition spseq : Type := nat -> prop.
\end{lstlisting}
Using this type, for instance the body of the sub-formula $\Enc_{\eqref{eq:pdr:f}}$ \eqref{eq:pdr:f:c} is specified as
\begin{lstlisting}[language=Coq, numbers=none, xleftmargin=10pt]
Fixpoint spseq_sseq (I : prop) (T : trans) 
(r : spseq) (ss : sseq) (i : nat) :=
  match i with
  | O => I ss.[0]
  | S i => spseq_sseq I T r ss i /\ 
                r i ss.[i] /\ T ss.[i] ss.[i+1]
  end.
\end{lstlisting}

\subsection{SMC as a Coq Proof Process}
\label{s:impl:smc}

In this section, we demonstrate that our specification works as a simple SMC tool (except for PDR as it requires to prepare an over-approximation).
To do so, a user first specifies a verification problem by defining $I$, $T$ and $P$ as objects of the types \verb|trans| and \verb|prop|, respectively.
Then, an SMC is performed by describing a theorem and its proof using a template script configured for each $\Enc_\bullet$.
Type checking of this proof with Coq mimics an execution of the SMC procedure Alg.~\ref{a:bmc}


For example, 
a template theorem and proof for SMC with \verb|forward_safety| (Eq.~\eqref{eq:forward}) is described as
%
\begin{lstlisting}[language=Coq, numbers=none, xleftmargin=10pt]
Theorem smc_example : forward_safety @$I$@ @$T$@ @$P$@ @$k$@. 
Proof.
  unfold @$I$@, @$T$@,@$P$@.
  unfold forward_safety; unfold ...
  repeat rewrite -> Nat.add_0_l; ...
  split.
  intros; smt solve; apply by_smt.
  repeat split; intros; smt solve; apply by_smt.
Qed.
\end{lstlisting}
%

Regarding Alg.~\ref{a:bmc}, an $\Enc_\bullet$ method is implemented as a Coq predicate (e.g., \verb|forward_safety| that implements $\Enc_{\eqref{eq:forward}}$), 
$\neg\mathsf{CheckSat}(\neg f)$ corresponds to discharging a sub-goal $f$ with \verb|smt solve|,
and $\mathsf{Decide}$ is represented as a proof template for each encoding method.

%

\begin{figure*}[!t]
\small
\begin{align*}
    \forall P,\ \forall s\!\in\! \texttt{sseq}, \forall i,j \!\in\! \mathbb{N},~ 
    & i \geq j \to P(s_i) \to P(\texttt{skipn}(i-j, s)_{j}). 
    & \text{(ss\&p 1)} \\
    %
    \forall T,\ \forall s\!\in\! \texttt{sseq}, \forall i,j \!\in\! \mathbb{N},~ 
    & \mathit{path}_T(s_{[j..(j+k)]}) \rightarrow \mathit{path}_T(\texttt{skipn}(j, s)_{[0..k]}). 
    & \text{(ss\&p 2)} \\
    %
    \forall T,\ \forall s\!\in\! \texttt{sseq}, \forall i,j \!\in\! \mathbb{N},~ 
    & \mathit{no\_loop}_T(s_{[j..(j+k)]}) \rightarrow \mathit{no\_loop}_T(\texttt{skipn}(j, s)_{[0..k]}).
    & \text{(ss\&p 3)} \\[1em]
    %
    \forall T,\ \forall s\!\in\! \texttt{sseq}, \forall j,k \!\in\! \mathbb{N},~
    & \mathit{path}_T(s_{[0..(j+k)]}) ~\leftrightarrow~ \mathit{path}_T(s_{[0..j]}) \land \mathit{path}_T(s_{[j..k]}).
    & \text{(ss\&p 4)} \\
    %
    \forall T,\ \forall s\!\in\! \texttt{sseq}, \forall j,k \!\in\! \mathbb{N},~
    & \mathit{loopF}_T(s_{[0..(j+k)]}) ~\rightarrow~ \mathit{loopF}_T(s_{[0..j]}) \land \mathit{loopF}_T(s_{[j..k]}).
    & \text{(ss\&p 5)} \\
    %
    \forall T,\ \forall s\!\in\! \texttt{sseq}, \forall i,j \!\in\! \mathbb{N},~
    & T(s_i, s_{i+1}) \land \mathit{path}_T(s_{[(i+1)..j]}) ~\leftrightarrow~ 
    \mathit{path}_T(s_{[i..j]}).
    & \text{(ss\&p 6)}
\end{align*}
\caption{Example lemmas on state sequences and paths.}
\label{f:ssp}
\end{figure*}

\section{Formal Verification}
\label{s:v}

We have verified the encoding methods in Sect.~\ref{s:smc} and \ref{s:smc:unbounded}.
The correctness (including soundness, completeness and termination) is discussed \Todo{rather informally} in the original papers\cite{Sheeran2000,Bradley2011};
our work aims at formalizing the soundness proofs with Coq, following the dissussions in \cite{Sheeran2000,Bradley2011}.
In this paper, we do not formalize the completeness and leave it as a future work
(see Sect.~\ref{s:concl}).
This section explains the process of formalization and proving.

Each encoding method represents a sufficient or necessary condition of the safety. 
Here, we describe the verified properties regarding Alg.~\ref{a:it}, which returns either $\True$ or $\False$ using two encoding methods.
In the following theorems, we relate the post-condition for true or false case with the safety (Eq.~\eqref{eq:safety}).
When the algorithm returns true using $\Enc_\bullet$, its soundness is described as follows.
\begin{theo}[Soundness of True Case with $\Enc_{\bullet}$] \label{th:sound:t}
    $\forall I, T, P, k,~ \Enc_\bullet(I,T,P,k) \ \rightarrow\  \mathit{safety}(I,T,P)$.
\end{theo}
Likewise, for the false case using $\Enc_\bullet$, its soundness is described as follows.
\begin{theo}[Soundness of False Case with $\Enc_{\bullet}$] \label{th:sound:f}
    $\forall I, T, P, k,~ \neg\Enc_\bullet(I,T,P,k) \ \rightarrow\  \neg\mathit{safety}(I,T,P)$.
\end{theo}
Note that Theorem~\ref{th:sound:f} does not ensure the completeness of the true case with $\Enc_\bullet$;
cf. $(\forall k, \neg q(k) \!\to\! \neg p)$ and $p \!\to\! (\exists k, q(k))$ are not logically equivalent.
%

Our verification is based on a shallow embedding of the transition systems and safety properties.
The above theorems, which relate the safety~\eqref{eq:ksafety} and the encoded formulas $\Enc_\bullet$, are specified directly in the logic of Coq.

In the end, we have provided formal proofs for each of the following combinations.
The proofs are explained in the following subsections.
\begin{itemize}
    \item {Theorem~\ref{th:sound:t}} with $\Enc_\eqref{eq:forward}$ and with $\Enc_\eqref{eq:backward}$: Sect.~\ref{s:v:fb}.
    \item {Theorem~\ref{th:sound:t}} with $\Enc_\eqref{eq:kinduction}$ ($k$-induction): Sect.~\ref{s:v:kind}.
    \item {Theorem~\ref{th:sound:f}} with $\Enc_\eqref{eq:ksafety}$: Sect.~\ref{s:v:bnd}.
    \item {Theorem~\ref{th:sound:t}} with $\Enc_\eqref{eq:pdr:t}$ (PDR): Sect.~\ref{s:v:pdr:t}.
    \item {Theorem~\ref{th:sound:f}} with $\Enc_\eqref{eq:pdr:f}$ (PDR): Sect.~\ref{s:v:pdr:f}.
\end{itemize}


In the proof of Theorem~\ref{th:sound:t} with $\Enc_{\text{\eqref{eq:forward}--\eqref{eq:kinduction}}}$, we restate the consequent safety into the form that considers only loop-free paths.
The following lemma is used for this deduction.
\begin{lemm} \label{th:loopfree}
    $\bigl( \forall i,\, s_{[0..i]},\, I(s_0) \to \mathit{loopF}_T(s_{[0..i]}) \to P(s_i) \bigr)$

    \noindent
    $\rightarrow \mathit{safety}(I,T,P)$.
\end{lemm}
Its proof is explained in Sect.~\ref{s:v:lf}.

\subsection{A Theory of State Sequences and Paths}
\label{s:v:ssp}

For the ease of proofs, we have developed vocabularies and lemmas on state sequences (of the type \verb|sseq|) and paths (\verb|sseq| values in which each pair of concatenated states satisfy $T$).
Some of the lemmas utilized in the verification are shown in Fig.~\ref{f:ssp}.

We introduce a suffix operation \verb|skipn| for state sequences, 
which is defined as $\texttt{skipn}(i, s)_{[m..n]} := s_{[(m+i)..(n+i)]}$,
and formalize the related properties.
Using \verb|skipn|, we have the lemmas such as Fig.~\ref{f:ssp} (ss\&p 1--3).

Various split operations for paths, loop-free paths, etc. are useful in the verification. Therefore, we formalize those split relations as lemmas; for instance, we have lemmas Fig.~\ref{f:ssp} (ss\&p 4--6).

\subsection{True Cases with $\Enc_{\eqref{eq:forward}}$ and with $\Enc_{\eqref{eq:backward}}$}
\label{s:v:fb}

\begin{figure}[!t]
    \centering
    \includegraphics[width=.45\textwidth]{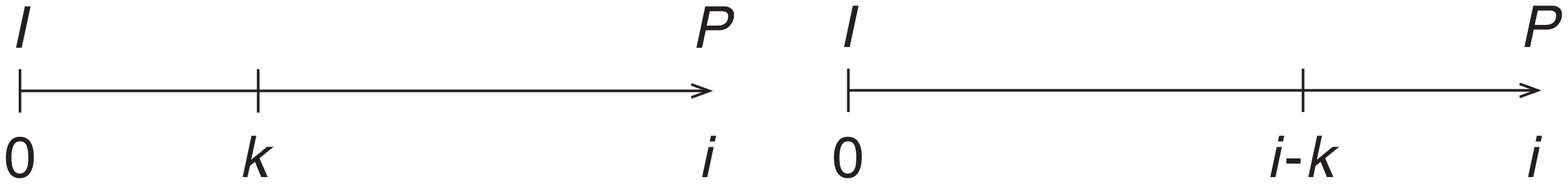} 
    \caption{Splitting of paths.}
    \label{f:split}
\end{figure}

Let $k$ be the parameter of the encoding method.
To prove Theorem~\ref{th:sound:t} for either $\Enc_\eqref{eq:forward}$ or $\Enc_\eqref{eq:backward}$,
we split either of the proof tasks into two cases $i \leq k$ and $i > k$, where $i$ is that in Lemma~\ref{th:loopfree}.
We prove that the hypothesis in Lemma~\ref{th:loopfree} holds for each case.

For the case $i \leq k$, we prove that the sub-formula $\Enc_\eqref{eq:ksafety}(I,T,P,k)$ implies the safety within the bound $k$. 
We perform an induction on $k$ to show the correspondence between $\Enc_\eqref{eq:ksafety}$, in which \verb|path| is specified recursively, and the consequent specified with a universal quantifier.

For the case $i > k$, 
we split the path constraint in the proof context in either way of the following:
\begin{gather}
    \label{eq:split:f}
    I(s_0) \land \mathit{loopF}_T(s_{[0..k]}) \land \mathit{loopF}_T(s_{[k..i]}) \to P(s_i), \\
    \hspace{-1em}
    \label{eq:split:b}
    I(s_0) \!\land\! \mathit{loopF}_T(s_{[0..(i\!-\!k)]}) \!\land\! \mathit{loopF}_T(s_{[(i\!-\!k)..i]}) \!\to\! P(s_i).
\end{gather}
The split paths are illustrated in Fig.~\ref{f:split}.
The first $\mathit{loopF}_T$ of Eq.~\eqref{eq:split:f} (resp. the second $\mathit{loopF}_T$ of Eq.~\eqref{eq:split:b}) matches with the last sub-formula of Eq.~\eqref{eq:forward} (resp. Eq.~\eqref{eq:backward}).
For Eq.~\ref{eq:split:b}, we utilize the \verb|skipn| operator in the unification of formulas ($s_{[(i-k)..i]}$ is rewritten as $\texttt{skipn}(i,s)_{[0..k]}$).
With the modifications of the proof contexts, the goals are discharged by matching them with the premises.

\subsection{True Case with $k$-Induction}
\label{s:v:kind}

Theorem~\ref{th:sound:t} with $\Enc_\eqref{eq:kinduction}$ is also proved with an application of Lemma~\ref{th:loopfree} and a case split into $i \leq k$ and $i > k$.

The case $i \leq k$ is proved as in Sect.~\ref{s:v:fb}.

For the case $i > k$, 
we apply a complete induction on the step number $i$ in the predicate $\mathit{safety}$;
we are to prove that the $i$-th state satisfies the property $P$ assuming $\Enc_\eqref{eq:kinduction}$ and the safety within the depth $i\!-\!1$.
The proof is done in three steps.
First, we split the constraint $\mathit{loopF}_T(s_{[0..i]})$ in the proof context in two, $\mathit{loopF}_T(s_{[0..(i-k)]})$ and $\mathit{loopF}_T(s_{[(i-k)..i]})$, as in Eq.~\eqref{eq:split:f}.
Second, we modify the indices in the sub-formula of Eq.~\eqref{eq:kinduction} with the lemma
\begin{align*}
    & \forall i, k,~ k < i \ \to\ \\
    & \bigl( \forall s_{[0..k]},\ \mathit{loopF}_T(s_{[0..k]}) \to \bigwedge{}_{0\leq i < k} P(s_i) \bigr) \to\\
    & \bigl( \forall s_{[(i-k)..i]},\ \mathit{loopF}_T(s_{[(i-k)..i]}) \to \bigwedge{}_{i-k \leq j < i} P(s_j) \bigr).
\end{align*}
Third, 
$P$ is inferred by applying the last sub-formula of Eq.~\eqref{eq:kinduction} and the hypothesis of the complete induction.

\subsection{False Case with $\Enc_{\eqref{eq:ksafety}}$}
\label{s:v:bnd}

Theorem~\ref{th:sound:f} with $\Enc_\eqref{eq:ksafety}$ is proved by showing that $\Enc_\eqref{eq:ksafety}$ always hold when assuming the safety of a system.
It is simply proved by an induction on $k\!+\!1$.

\subsection{True Case with PDR}
\label{s:v:pdr:t}

Theorem~\ref{th:sound:t} with $\Enc_\eqref{eq:pdr:t}$ is proved by several case analyses and inductions as described below.
At first, the proof goal is split into the two goals of the form
\begin{multline}
\label{eq:pdr:bnd}
    \forall I, T, P, k,~ \Enc_\eqref{eq:pdr:t}(I,T,P,k) \rightarrow \\
    \forall i, i \circ k\!+\!1 \rightarrow \forall s_{[0..i]}, I(s_0) \!\to\! \mathit{path}(s_{[0..i]}) \!\to\! R_i(s_i),
\end{multline}
where $\circ$ is set as either $\leq$ or $>$.
The consequence $R_i(s_i)$ is obtained by applying $\Enc_\eqref{eq:pdr:t}$ (b) to the term $P(s_i)$ in $\mathit{safety}$.

For the case $\circ :=\ \leq$, we perform an induction on $i$.
The initial case is proved by $\Enc_\eqref{eq:pdr:t}$ \eqref{eq:pdr:t:a}.
In the induction step where $i\!+\!1 \leq k\!+\!1$,
the proof context consists of the consequence $R_{i+1}(s_{i+1})$ and a set of hypotheses.
$R_{i+1}(s_{i+1})$ is transformed to $R_{i}(s_{i})$ by $\Enc_\eqref{eq:pdr:t}$ (c).
As a result, the proof context matches the induction hypothesis and be discharged.


For Eq.~\eqref{eq:pdr:bnd} with $\circ :=\ >$, 
we again perform an induction on $i$.
The initial case ($0 > k'$, where $k := k\!+\!1$) immediately holds.
In the induction step ($i+1 > k'$),
the goal $R_{k'}(s_{i+1})$ is transformed into $R_{k'}(\texttt{skipn}(i, s)_1)$.
Then, we have $R_k(\texttt{skipn}(i, s)_0) = R_k(s_{i})$ by applying $\Enc_\eqref{eq:pdr:t}$ (d), and $R_{k'}(s_{i})$ by rewriting with $\Enc_\eqref{eq:pdr:t}$ (e).
Here, we consider two cases $i = k'$ and $i > k'$ separately.
For the first case, the goal $R_{k'}(s_{i}) = R_{k'}(s_{k'})$ is proved by applying Eq.~\eqref{eq:pdr:bnd} with $i := k'$.
For the case when $i > k'$, 
we are able to deduce the goal $R_{k'}(s_i)$ from the induction hypothesis; so it is discharged.

\subsection{False Case with PDR}
\label{s:v:pdr:f}

Theorem~\ref{th:sound:f} with $\Enc_\eqref{eq:pdr:f}$ is proved by showing that each of the three sub-formulas of $\Enc_\eqref{eq:pdr:f}$ holds when assuming the safety of a system.

The sub-formulas (a) and (b) are easy to prove.

Before the proof of the sub-formula $\Enc_\eqref{eq:pdr:f}$ (c), we prepare the following lemma that states a necessary condition.
\begin{multline}
    \label{eq:pdr:neg}
    \forall I, T, P, R_{[0..(k-1)]}, s_{[0..k]},~ \\
    \Bigl( I(s_0) \land \bigwedge_{0 \leq i < k} ( R_i(s_i) \land T(s_i,s_{i+1}) ) \land \neg P(s_{i+1}) \Bigr)
    \to \\
    I(s_0) ~\land~
    \bigl( \forall i, i < k \rightarrow \mathit{path}_T(s_{[0..(i+1)]}) \land R_i(s_i) \bigr).
\end{multline}
Note that the hypothesis is the negation of the body of $\Enc_\eqref{eq:pdr:f}$ (c).
It is proved by an induction on $k$; in the induction step, we prove by a case split into the cases 
$k = 0$, $0 = i < k$ and $0 < k \leq i+1$.

The proof context of the sub-formula $\Enc_\eqref{eq:pdr:f}$ (c) contains the hypothesis part of Eq.~\eqref{eq:pdr:neg}, so it is rewritten into the consequence part of Eq.~\eqref{eq:pdr:neg}.
The proof goal $P(s_{k+1})$ is discharged by the assumption of the safety, but we need to show additionally that $s_{[0..(k+1)]}$ is a path of the system.
Using the rewritten premise above, we are able to prove it.

\subsection{Reduction to Loop Free Paths}
\label{s:v:lf}

\begin{figure}[!t]
    \centering
    \vspace*{-1.5em}
    \subfloat { \includegraphics[width=.23\textwidth]{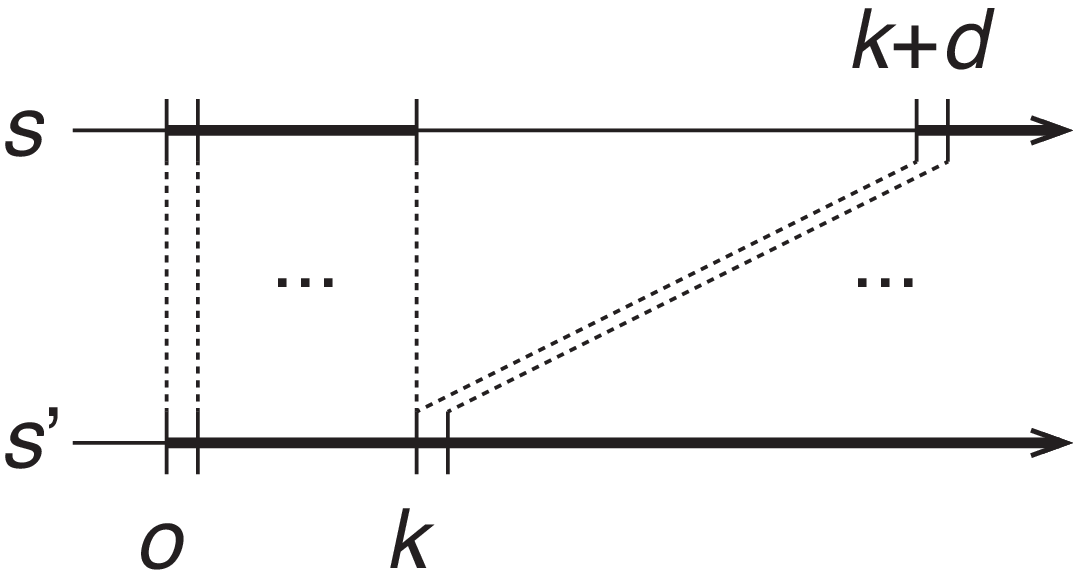} }
    \subfloat { \includegraphics[width=.23\textwidth]{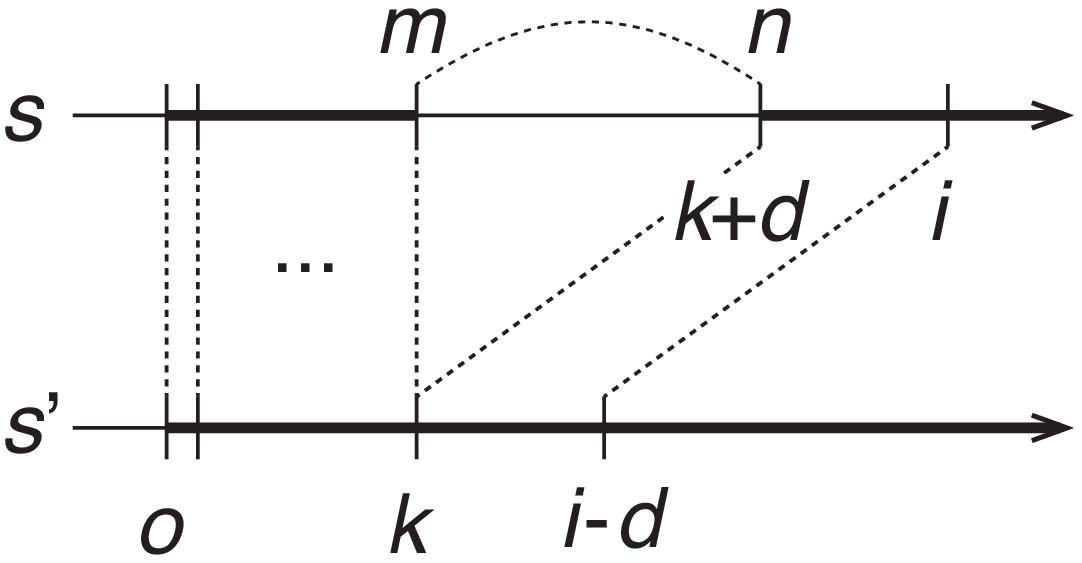} }
    \caption{\texttt{shorter\_ss} (left) and the lemma~\eqref{eq:shorter} (right).}
    \label{f:shorter}
\end{figure}

Lemma~\ref{th:loopfree} is proved by a complete induction on $i$ in $\mathit{safety}$.
In the induction step, we have to show the safety of the arbitrary paths of length $i\!+\!1$, assuming the safety within the lengths less than $i\!+\!1$ and the safety of the loop-free paths of length $i\!+\!1$.
It is proved based on the observation that the reachable states by looped paths can be reached by a shorter path.

To compare paths with or without a loop reaching a same state, we introduce the predicate $\texttt{shorter\char`_ss}(o, k, d, s, s')$ defined as
\begin{equation*}
    \forall i \!\in\! \mathbb{N},~
    \bigl( (i \leq k-o \rightarrow s_{k-i} = s'_{k-i}) ~\land~
    s_{k+d+i} = s'_{k+i} \bigr).
\end{equation*}
Fig.~\ref{f:shorter} (left) illustrates the definition.
It represents the fact that a sequence $s'$ is a shortened sequence of $s$ where the fragment $s_{[(k+1)..(k+d-1)]}$ is removed from $s$;
thus, two fragments of each of the two sequences coincide as $s_{[o..k]} = s'_{[o..k]}$ and $s_{[(k+d)..]} = s'_{[k..]}$.

With this predicate, we formalize the lemma
\begin{multline} \label{eq:shorter}
    \forall i \!\in\! \mathbb{N},\ \forall s \!\in\! \texttt{sseq},
    \Bigl( \bigvee_{0 \leq m < n \leq i} s_m = s_n \Bigr) ~\rightarrow~ \\
    \neg\bigl(\ \forall s' \!\in\! \texttt{sseq},~ \forall k,d \!\in\! \mathbb{N},~~
    k+d > i \ \lor\ d \leq 0 \ \lor\ \\
    \texttt{shorter\char`_ss}(0, k, d, s, s') \ \land\ s_i \neq s'_{i-d}\ \bigl).
\end{multline}
Note that the consequent part can be transformed into the formula below in classical logic.
\begin{multline*} \label{eq:shorter:neg}
    \exists s' \!\in\! \texttt{sseq},~ \exists k,d \!\in\! \mathbb{N},~~
    k+d \leq i \ \land\ d > 0 \ \land\ \\
    \texttt{shorter\char`_ss}(0, k, d, s, s') \ \land\ s_i = s'_{i-d}
\end{multline*}
Fig.~\ref{f:shorter} (right) illustrates this lemma.
It states that, when a sequence $s$ contains a loop, there exists a shortened sequence $s'$ and the same states are reachable with the two paths.
The consequent holds by matching the variables as $k = m$ and $d = n-m$.


To prove Lemma~\ref{th:loopfree}, we perform a proof by contradiction; we assume that an unsafe path of length $k$ exists when the safety is checked for loop-free paths of length $k$.
Such path should be looped; thus, we can show that there are unsafe shorter paths, which contradicts the induction hypothesis.

\section{Review on the Verification Result}
\label{s:res}

This section describes the statistical data (Sect.~\ref{s:res:stat}) and discussions (Sect.~\ref{s:res:discuss} and \ref{s:res:cl}) regarding the overall results of the verification with Coq described in Sect.~\ref{s:v}.

\subsection{Statistics}
\label{s:res:stat}

\begin{table}[!t]
    \centering
    \caption{LOCs of the proof scripts.}
    \label{tbl:loc}
    \begin{tabular}{l|rr||l|rr} \hline
        & total & ss\&p &
        & total & ss\&p \\
        \hline
        Th.1 w. $\Enc_\eqref{eq:forward}$    & 459 & 348 &
        Th.2 w. $\Enc_\eqref{eq:ksafety}$    &  14 &   0 \\
        Th.1 w. $\Enc_\eqref{eq:backward}$   & 545 & 423 &
        Th.1 w. $\Enc_\eqref{eq:pdr:t}$      & 147 &  70 \\
        Th.1 w. $\Enc_\eqref{eq:kinduction}$ & 607 & 423 &
        Th.2 w. $\Enc_\eqref{eq:pdr:f}$      &  74 &   0 \\
        Lemma~\ref{th:loopfree}                & 352 & 270 &
        && \\
        \hline
    \end{tabular}
\end{table}

The LOC for each proof is shown in Table~\ref{tbl:loc}; the third and sixth columns show the LOC of the proofs related to state sequences and paths described in Sect.~\ref{s:v:ssp} and \ref{s:v:lf}.
For instance, the overall proof script for Theorem~\ref{th:sound:t} with $\Enc_\eqref{eq:forward}$ consisted of 459~LOC, which involved the proof script for Lemma~\ref{th:loopfree}, three lemmas for the cases analyzed and 11~``ss\&p''~lemmas. 
%
Note that, many of the ``ss\&p'' lemmas were shared between the different verification tasks.
%

The proof of Lemma~\ref{th:loopfree} was formalized with 352~LOC,
in which we reduced the lemma twice into other proof goals; the resulting goal was proved with nine ``ss\&p'' lemmas.
%

\subsection{Discussions}
\label{s:res:discuss}

We have successfully formalized the soundness proofs following the discussions in the original papers \cite{Sheeran2000,Bradley2011}.
In the formalization, 
we used the lemmas in Sect.~\ref{s:v:ssp} and \ref{s:v:lf} and they were reused in several proof tasks.

In the proofs, a number of inductions were performed based on the recursive definitions of the encoding predicates and the path lengths under consideration.
In each induction step, proof context was carefully modified to apply prepared lemmas;
in some proofs, rewriting with the \verb|skipn| expression and applying path splitting lemmas were useful.

A fair amount of LOCs were devoted to Lemma~\ref{th:loopfree} as it required to perform a proof by contradiction and translation between loop paths and shorter paths (Sect.~\ref{s:v:lf}).

The proofs for PDR (Sect.~\ref{s:v:pdr:t} and \ref{s:v:pdr:f}) consisted of a number of case analyses and inductions; the split goals were discharged by applying sub-formulas of $\Enc_\eqref{eq:pdr:t}$ and $\Enc_\eqref{eq:pdr:f}$.
It was less often to describe paths explicitly in the proof contexts than the other methods and thus the resulting proofs became relatively small.

\subsection{Use of Classical Logic}
\label{s:res:cl}

In the proof, we used several lemmas of the \texttt{Classical\_prop} module.
On the other hand, we tried to minimize the proofs that require those lemmas.
As a result, we identified the two parts in the proofs:
(i) the transformation into conjunctions at the front-end of each encoding method (e.g. from \eqref{eq:ksafety} to \eqref{eq:ksafety:cnf}, from \eqref{eq:forward} to \eqref{eq:forward:cnf}, etc.).
(ii) the proof of Lemma~\ref{th:loopfree}.
The other proofs were formalized without the use of classical logic.

We conjecture that the use of the law of excluded middle is essential in the parts (i) and (ii).
In (ii), it was used to apply the law of double negation to perform a proof by contradiction, 
to obtain the contraposition of Eq.~\eqref{eq:shorter}, and
to decompose the negation of the \verb|loop_free| clause.

\section{Related Work}
\label{s:related}

As far as we know, there has been no formally verified SMC tool.
Issues in the SMC tools include the possibility of flaws in encoding methods and their implementations.
Our tool 
demonstrates that a reliable SMC is possible by directly using a verified implementation.
%
On the other hand, lack of certification for (especially ``safe'') verification results becomes another issue~\cite{Namjoshi2001,Mebsout2016}.

For generic model checking, there exist several verified tools.
Sprenger~\cite{Sprenger1998} has formalized the modal $\mu$-calculus and a dedicated model checker in Coq.
Amjad~\cite{Amjad2003} has proposed to embed a symbolic model checker with its underlying BDD within the HOL tool.
More recently, Esparza et al.~\cite{Esparza2013} have developed more practical and verified LTL model checker, which has been specified and verified in Isabelle/HOL and then extracted as an ML implementation.
%
Wimmer and Lammich~\cite{Wimmer2018} have proposed a verified model checker based on timed automata, in which model checking with abstraction of continuous states has been formalized and verified.
In those verification tasks, the correctness of model checking algorithms was verified by relating the outputs of algorithms and the semantics of properties.
When compared to our work, formalization with state sequences and paths that considers lengths/shifting/splitting seems specific in our work.
In the previous work except \cite{Esparza2013}, the formalizations were based on the set of states and paths were not considered explicitly.
In \cite{Esparza2013}, paths were considered but operations seemed simpler than ours.

As a slightly different line of work, there have been methods that generate proof certifications from provers e.g. SAT/SMT solvers~\cite{Armand2011,Mebsout2016} and model checkers~\cite{Namjoshi2001}. They propose to verify a result computed by the provers by verifying the generated certificates on a theorem prover.
Our purpose is different from those works but integration of our work with certification might enable constructive combined SMC and deduction.

\section{Conclusions}
\label{s:concl}

We specified the SMC methods and formalized their soundness proofs on Coq.
We consider that our result provides an example formal proof, which is not trivial to perform on a proof assistant like Coq.
The specification of the SMC methods and the soundness proofs are available at \url{https://github.com/dsksh/coq-smc/}.

There are several future work directions.
First, we can continue the verification task to obtain the formal correctness proof.
A difficulty in the proof of completeness will be the formalization of properties
such as ``an infinite sequence of finite states should contain a loop.''
We consider that it requires another effort in the development of the theory of paths.
Second direction is to consider other SMC methods (including newly improved methods), target systems and properties,
e.g. liveness properties and the $k$-Liveness method (e.g. \cite{Claessen2012}).
Otherwise, formalization of the encoding methods based on the bounded semantics~\cite{Biere1999} 
will be interesting.
Third, 
%
a verification and extraction of a practical SMC tool 
will be valuable.


\vspace{1em}

\emph{Acknowledgment}.
This work was partially funded by JSPS (KAKENHI 18K11240).

\bibliographystyle{IEEEtran}
\bibliography{article}

\end{document}